\documentclass[preprint,12pt]{elsarticle}
\usepackage{graphicx}
\usepackage{caption}
\usepackage{subcaption}
\usepackage{amsmath, latexsym, soul, color, type1cm,dsfont, float, fancyhdr}

\usepackage{graphicx}
\usepackage{amssymb}

\usepackage{lineno}

\journal{Journal of Magnetism and Magnetic Materials}

\usepackage[normalem]{ulem}

\begin{document}

\begin{frontmatter}


\title{Scaling the effect of the dipolar interactions on the ZFC/FC curves of random nanoparticle assemblies}



\author[USC]{David Serantes\corref{mycorrespondingauthor}}
\cortext[mycorrespondingauthor]{Corresponding author}
\ead{david.serantes@usc.es}
\address[USC]{Applied Physics Department and Instituto de Investigaci\'{o}ns Tecnol\'{o}xicas,
  Universidade de Santiago de Compostela, E-15782 Campus Vida s/n,
  Santiago de Compostela, Spain}

\author[UU]{Manuel Pereiro}
\address[UU]{Department of Physics and Astronomy, Uppsala University, P.O. Box 516, 751 20 Uppsala,
Sweden}

\author[UoY]{Roy Chantrell}
\address[UoY]{Department of Physics, University of York, York, YO10 5DD, United Kingdom}

\author[USC]{Daniel Baldomir}

\begin{abstract}
\textit{Zero field cooling} (ZFC) and \textit{field cooling} (FC) protocols are commonly used to investigate the properties of magnetic nanoparticle systems. For non-interacting conditions the particle properties are fairly well correlated with the shape of the ZFC/FC curves. However, that is not the case when significant dipolar interparticle interactions (DII) are present, what frequently occurs in experimental samples (e.g. aggregates in biological systems; or the dried powder often used for the ZFC/FC measurements). The purpose of this work is to show how the influence of the DII on the ZFC/FC curves, computed by the volume sample concentration $c$, can be described in a general way if scaled by the dimensionless parameter $c_{0}=2K/M_{S}^{2}$; where $K$ and $M_{S}$ are the anisotropy and saturation magnetization constants of the particles, respectively. This scaling parameter, which is straightforwardly derived from the energy equation governing the system, has an analogous meaning to the normalization of the external magnetic field $H$ by the anisotropy field of the particles $H_{A}=2K/M_{S}$. We use a Monte Carlo technique to show how apparently different $T_{B}$ \textit{vs.} $c$ curves of various particles types (where $T_{B}$ is the \textit{blocking temperature}), follow the same trend if scaling $c/c_{0}$.
\end{abstract}


\end{frontmatter}


\section{Introduction}
\label{S:1}

Magnetic nanoparticles are receiving increasing attention in the last years particularly based on their promising nanomedical uses,\cite{pelaz_acsnano2017,colombo_csr2012} what requires detailed understanding of their magnetic properties. A major difficulty for their \textit{in vivo} application is that the particles are usually randomly arranged when embedded in the biological matrices, and aggregation after cell internalization is very frequent\cite{rojas_act.biomat.2017}. The problem, regarding the property control, is that for those closely-arranged systems the dominant interaction is the magnetic dipole-dipole coupling, which lacks an analytical solution and exhibits a complex interplay with the other more relevant energies in the system (anisotropy\cite{serantes_prb2012}, Zeeman\cite{serantes_jncs2008}). Finding a precise and easy description of its features arises, therefore, as a crucial problem to address in the current research efforts towards the nanomedical revolution.

Such a purpose has often been pursued on the basis of the N\'eel model: the dipolar interaction effects are investigated in terms of their effects on the anisotropy energy barrier of single-domain \textit{superparamagnetic} (SPM) systems and their corresponding effect on the characteristic \textit{blocking temperature}, $T_{B}$ \cite{kechrakos_review2010}. The interparticle coupling strength is varied by changing the magnetic particle concentration of the system, $c$. The reasons to choose this method are the ease of carrying out the experimental measurements and to correlate them with the theoretical background \cite{vargas_prb2005}.

Despite the intense research devoted to this field in recent years, no conclusive  understanding of the effects of interactions on the blocking temperature have been yet achieved: one can find in the literature very different $T_{B}(c)$ trends for essentially very similar samples. These trends range from, i) a regular increase \cite{tomita_prb2007}; ii) a rapid increase followed by a saturation \cite{vestal_jpcB2004}; iii) a non-monotonic behaviour \cite{serantes_prb2010}; iv) a decrease \cite{hansen_jmmm1998}; etc. Although analytical models have been developed that describe reasonably well specific cases, a general model able to account for all the different shapes has not been reported so far.

In an effort to shed light on this complex scenario, we present in this work an alternative approach: by simply rewriting the energies governing the system in an appropriate way, the influence of the dipolar interactions on the magnetic behavior of the system (introduced as proportional to $c$) is described in a general way by the unitless parameter $c_{0}=2K/M^{2}_{S}$. It must be noted that the $2K/M^{2}_{S}$ ratio is well known to provide the relation of the anisotropy field to the maximal dipolar interaction field\cite{berkov_jpcm2006}, and has been extensively used particularly to address the competition between anisotropy and shape effects\cite{kharel_jpd2013,usov_jap2012,roy_jap2014}. In the case of interacting nanoparticles, ourselves\cite{martinez-boubeta_afm2012,ruta_sr2015} and others\cite{usov_aipadv2016} have routinely (and successfully) applied it to explain the effects of interactions through hysteresis properties. However, to the best of our knowledge its general character has not been explained in detail so far, neither has it been applied to the specific case of the diversity of experimental trends obtained through ZFC/FC measurements, except in limited specific cases \cite{serantes_prb2010,hoppe_jpcc2008}. Here we investigate the underlying thermodynamic basis of the scaling law. In order to demonstrate the universality of this scaling parameter, we build up our arguments in the context of the well-known usual single-domain particles SPM framework. We demonstrate numerically, using a Monte Carlo model, that the scaling applies for a system of randomly spatially distributed nanoparticles. The scaling approach is proposed as an interesting basis for the investigation of experimental measurements; deviations from the scaling law could arise, for example, from clustering effects.

\section{Scaling the governing energies of the system}

The SPM model represents the particles by their large magnetic \textit{supermoment} and also by their magnetic anisotropy $\vec{K}=K\hat{n}_{i}$, with K the anisotropy constant and $\hat{n}_{i}$ the easy axis direction. The magnetic supermoment results from the coherent rotation of the inner atomic moments and is to first approximation proportional to the particle volume $V_{i}$ as $\vec{\mu}_{i}=M_{S}V_{i}\hat{e}_{{\mu}_{i}}$, with $M_{S}$ the saturation magnetization and $\hat{e}_{{\mu}_{i}}$ the unitary vector that describes its orientation.

The energy per particle $E^{(i)}$ for a non-interacting system is given by the anisotropy ($E_{A}$) and Zeeman ($E_{Z}$) energies as
\begin{equation}\label{Ei_non-interact}
	E^{(i)}=E^{(i)}_{A}+E^{(i)}_{Z}=-KV_{i}\left({\frac{\vec{\mu}_{i}\cdot\hat{n}_{i}}{\left|\vec{\mu}_{i}\right|}}\right)^{2}-\vec{\mu}_{i}\cdot{\vec{H}}
\end{equation}

Both energy terms determine the magnetic evolution of the system. However, although the intrinsic thermal stability (characterised by the zero-field blocking temperature $T_B=KV/25k_B$), the field dependence is determined by the ratio of the anisotropy and Zeeman terms. This is easily emphasized on rewriting Eq.~(\ref{Ei_non-interact}) as 
\begin{equation}\label{ei_non-interact}
	e^{(i)}=\frac{E^{(i)}}{2KV_{i}}=-\frac{1}{2}\cos^{2}{\varphi_{i}}-\frac{H}{H_{A}}\cos{\theta_{i}}
\end{equation}
where $\varphi_{i}$ and $\theta_{i}$ are the angles that $\vec{\mu}_{i}$ forms with respect to $\vec{H}$ and $\vec{K}$, respectively. Likewise, the \textit{anisotropy field} is defined as $H_{A}=2K/M_{S}$, which accounts for the relative importance of the Zeeman and anisotropy energies. This is highlighted in that, for a system with randomly oriented easy axes, the field dependence of $T_{B}\propto{(1-H/H_{A})^{3/2}}$ \cite{victora_prl1989}. This expression illustrates that the influence of $H$ on the evolution of the magnetisation of a nanoparticle system does not depend on the absolute $H$-value but rather on the ratio $H/H_{A}$. 
Thus, particles with different values of $K$ and $M_{S}$ may still follow a similar $T_{B}(H)$ shape depending on their characteristic $H_{A}$ value, i.e. $T_{B}\equiv{T_{B}(H/H_{A})}$.

We follow the same line of reasoning to investigate the influence of the dipolar interaction. In this case, for an $N$-particle system the energy per particle is
\begin{eqnarray}\label{Ei_interact}
\lefteqn{E^{(i)}=-KV_{i}\left(\frac{\vec{\mu}_{i}\cdot\hat{n}_{i}}{\left|\vec{\mu}_{i}\right|}\right)^{2}-\vec{\mu}_{i}\cdot\vec{H}+} & & \nonumber
\\& & +\sum^{N}_{j\neq{i}}\left(\frac{\vec{\mu}_{i}\cdot\vec{\mu}_{j}}{r^{3}_{ij}}-3\frac{(\vec{\mu}_{i}\cdot\vec{r}_{ij})(\vec{\mu}_{j}\cdot\vec{r}_{ij})}{r^{5}_{ij}}\right)
\end{eqnarray}where $\vec{r}_{ij}$ is the vector connecting particles $i$ and $j$. Our procedure is to rewrite Eq.~(\ref{Ei_interact}) also in terms of $H/H_{A}$, trying to find also an easy dependence for the dipolar energy term. For such a purpose, we assume i) a monodisperse system; ii) the particles are located into a cubic box of side $L$, so that we may use normalized units $\vec{a}_{ij}=\vec{r}_{ij}N^{\frac{1}{3}}/L$; iii) the dimensionless sample concentration $c$ results in $c=\sum^{N}_{i=1}V_{i}/L^{3}\equiv{NV/L^{3}}$. Thus, the reduced energy $e^{(i)}$ reads now
\begin{eqnarray}\label{ei_reducidas}
\lefteqn{e^{(i)}=\frac{E^{(i)}}{2KV}=-\frac{1}{2}\cos^{2}\varphi_{i}-\frac{H}{H_{A}}\cos\theta_{i}+} & & \nonumber
\\& & +\frac{c}{c_{0}}\sum^{N}_{j\neq{i}}\left(\frac{\hat{e}_{{\mu}_{i}}\cdot\hat{e}_{{\mu}_{j}}}{a^{3}_{ij}}-3\frac{(\hat{e}_{{\mu}_{i}}\cdot\vec{a}_{ij})(\hat{e}_{{\mu}_{j}}\cdot\vec{a}_{ij})}{a^{5}_{ij}}\right) 
\end{eqnarray}where we have introduced $c_{0}=2K/M^{2}_{S}$. The $c_{0}$-parameter weights the importance of the dipolar interaction energy as $c/c_{0}$, analogously as $H/H_{A}$ weights the importance of the Zeeman energy. Therefore, we suggest, similarly to the fact that $H/H_{A}$ provides a general scaling behaviour for the magnetic field-dependence of the system, it is possible to define also a scaling law for the influence of the dipolar interaction, given by $c/c_{0}$. This result demonstrates the existence of an inner scale of energies in the thermomagnetic properties of those nanostructured systems as a function of the $H_{A}$ and $c_{0}$ parameters.

The physical meaning of $c_{0}$ can be interpreted as a characterisation of the relative importance between the anisotropy and dipolar energies, i.e. $c_{0}=2K/M^{2}_{S}\propto{E_{A}/E_{D}}$. It is worth to emphasize again that, although the $2K/M^{2}_{S}$ ratio has been proposed in the literature as a relevant quantity concerning interacting SPM systems (see e.g. Refs. \cite{garcia-otero_prl2000,berkov_jmmm2001,jonsson_acp2004}), to the best of our knowledge neither its interpretation as a general scaling factor nor its potential to resolve controversial results in the literature (as the $T_{B}$ \textit{vs.} $c$ evolution) have previously been reported.

\section{Monte Carlo simulations}

The next step is to demonstrate that the above assertion on $c_{0}$ as a scaling factor for the influence of magnetic dipolar interactions can effectively solve controversial results in the literature. With this purpose in mind, we decided to tackle the $T_{B}$ \textit{vs.} $c$ problem based on its central role amongst theoretical-nanoparticle problems as to dipolar interaction effects.

Our procedure consists of comparing the $T_{B}(c)$ data for different types of particles, in order to study the different trends reported in the literature. We represent the particle types by their corresponding $c_{0}$-values, and examine their magnetic behaviour both as a function of $c$ and $c/c_{0}$. Since experimental data is usually subjected to uncontrolled secondary effects (polydispersity \cite{wang_apl2009}, aggregation \cite{hoppe_jpcc2008}) that might mask the influence of the dipolar interactions, we have used a Monte Carlo technique that allowed us to perfectly control the characteristics of the sample and hence to be sure about the nature of the different $T_{B}$ data. The computational procedure is the same as described in Ref. \cite{serantes_prb2010}, and the results are presented in the usual reduced temperature units $t=k_{B}T/2KV$. $T_{B}$ is roughly evaluated as the maximum of the \textit{zero field cooling} (ZFC) curves under a low applied field $H=100~Oe$, as usual, and hence to obtain the $T_{B}$ \textit{vs.} $c$ data we simulated ZFC processes for a systematic variation of $\Delta{c}$ in steps of $2.5\%$. It is important to be aware that $T_{B}$ does not correspond exactly to the peak of the ZFC curve, but has a smaller value (see e.g. I. J. Bruvera \textit{et al.} \cite{bruvera_jap2015} for an insightful discussion); the choice of the peak was made for the sake of simplicity, since our arguments are applicable to all features of the ZFC/FC curves.

\begin{table}[!ht]
\caption{$M_{S}(emu/cm^{3})$ data and the corresponding $c_{0}$, $H_{A}(Oe)$, and $h_{100~Oe}=100~Oe/H_{A}$ values for $K=1.5\times10^{5}erg/cm^{3}$.}
\begin{center}
\begin{tabular}{c c c c c  c}
\hline
\hline
$M_{S}$ & & $c_{0}$ & & $H_{A}$ & $h_{100~Oe}$ \\
\hline
274 & & 4.00 & & 1095 & 0.09 \\
316 & & 3.00 & & 949 & 0.10 \\
387 & & 2.00 & & 775 & 0.13 \\
548 & & 1.00 & & 548 & 0.18 \\
775 & & 0.50 & & 387 & 0.26 \\
949 & & 0.33 & & 316 & 0.32 \\
1095 & & 0.25 & & 274 & 0.36 \\
\hline
\hline
\end{tabular}
\end{center} 
\label{table-c0}
\end{table}

To represent different types of particles we characterized them by their $K$ and $M_{S}$ values, and associated each type to the corresponding $c_{0}$. To select the $K$ and $M_{S}$ values (and hence the $c_{0}$ cases) we decided to keep a common value of $K$ and vary $M_{S}$. By proceeding in this way we can expect that in the non-interacting limit case $T_{B}$ will be the same for the different particle types, since $T_{B}\approx{KV/25k_{B}}$ (i.e. independent of $M_{S}$), so that we can study the evolution of $T_{B}$ with $c$ from this common non-interacting point. It is worth to note that although this is an ideal assumption, it is in fact possible to design different materials so that they have a common (or very similar) $K$ value but different $M_{S}$. Also, for computational purposes it is easier to keep the same value of $K$ (included in the reduced temperature units) and vary only $M_{S}$. The $K$ value is relatively low so that dipolar interactions influence easily the system, using $K=1.5\times10^{5}erg/cm^{3}$. The $M_{S}$ values considered and the corresponding $H_{A}$ and $c_{0}$ parameters are shown in Table \ref{table-c0}, together with $H/H_{A}$ ratio corresponding to $H=100~Oe$.

The $M_{S}$ values were chosen so that they represent realistic physical values, and the equivalent $c_{0}$ cases cover a wide range of $E_{A}$ \textit{vs.} $E_{D}$, so that $c_{0}$ varies between the $E_{A}$-dominating case ($c_{0}=4$), to the $E_{D}$-dominating one ($c_{0}=1/4$), covering as well intermediate cases. In Fig. \ref{normaliza_h,c0}, we show some ZFC curves for specific field and concentration conditions in order to illustrate the equivalent role of $c_{0}$ as a scaling parameter for the sample concentration analogous to the $H_{A}$ value for the magnetic field.

\begin{figure}[!ht]
\includegraphics[angle = 0,width = 1.0\textwidth]{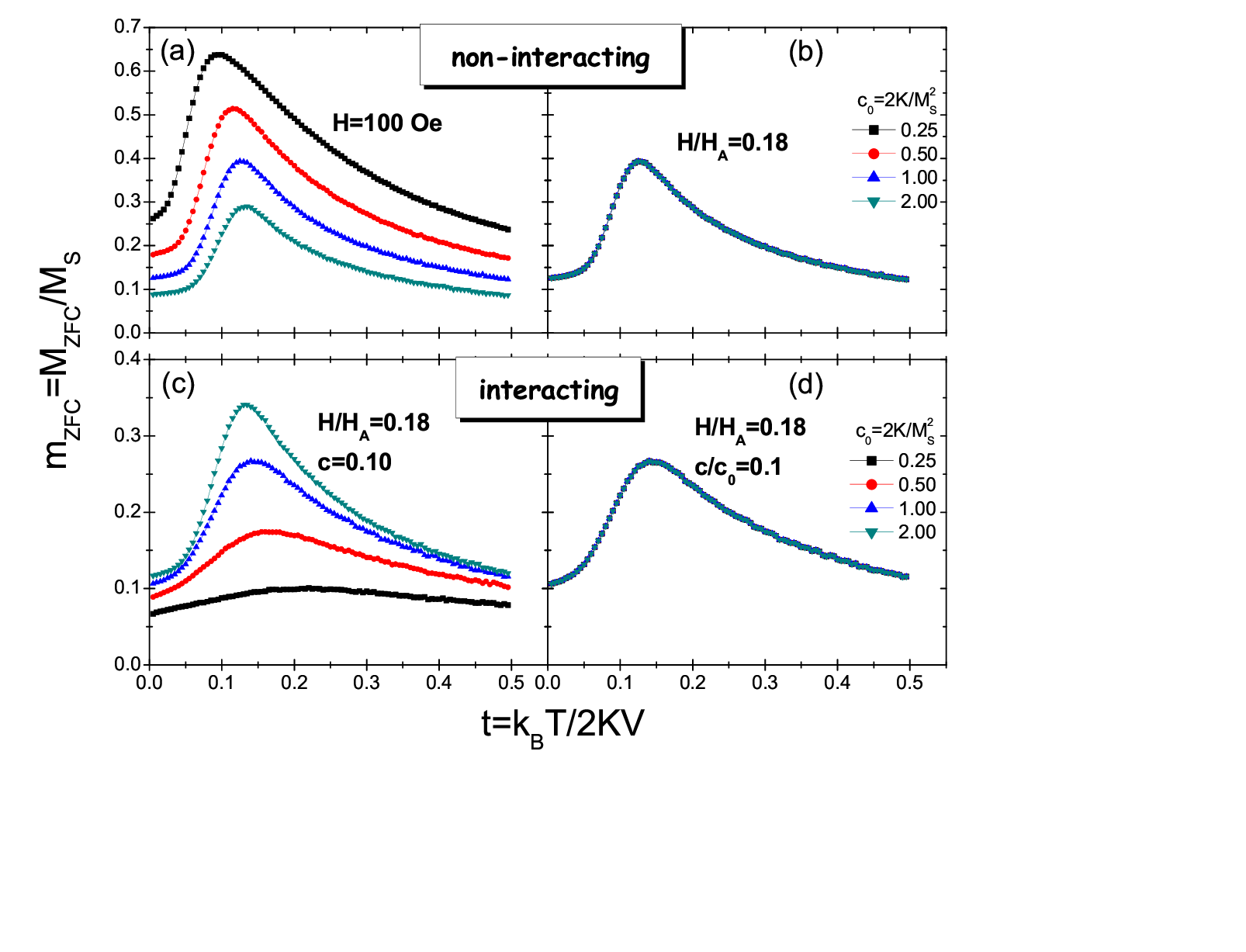}
\caption{(Color online) ZFC curves for different field and interaction conditions, for some $c_{0}$ values: $H=100~Oe$, $c=0.0$ in (a); $H/H_{A}=0.18$, $c/c_{0}=0.0$ in (b); $H/H_{A}=0.18$, $c=0.10$ in (c); $H/H_{A}=0.18$, $c/c_{0}=0.10$ in (d).}\label{normaliza_h,c0}
\end{figure}

Figure \ref{normaliza_h,c0}(a) displays some ZFC curves for the non-interacting case under the same absolute field $H=100~Oe$. In this case, the curves differ in form and have different maxima; however, if using the same reduced field $H/H_{A}=0.18$ we see in Fig. \ref{normaliza_h,c0}(b) that all curves overlap and result therefore in the same $T_{B}$ value, as expected. However, for interacting conditions and under the same $H/H_{A}=0.18$ value, the ZFC curves separate if using the same $c=0.10$ value and exhibit different $T_{B}$ values; however, if using the same $c/c_{0}=0.1$ value the curves overlap and $T_{B}$ is constant. This demonstrates the suitability of $c_{0}$ as a scaling factor for the dipolar interaction analogous to $H_{A}$ for the magnetic field.

\begin{figure}[!ht]
\includegraphics[angle = 0,width = 1.0\textwidth]{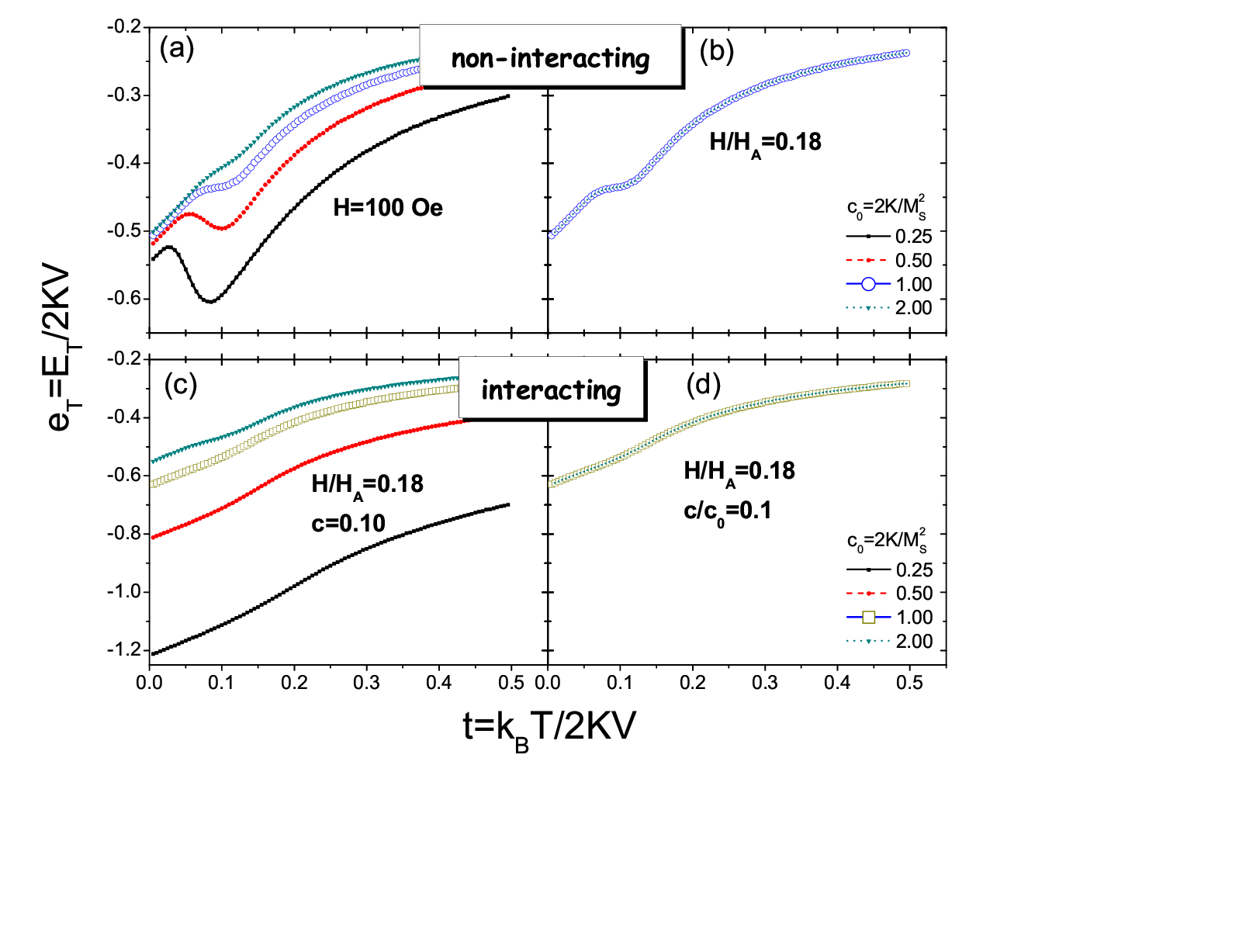}
\caption{(Color online) $e_{T}=E_{T}/2KV$ data corresponding to the ZFC curves displayed in Fig. \ref{normaliza_h,c0}.}\label{comparando_c0}
\end{figure}

To illustrate the inner scale of energies as the origin of this scaling behavior, we show in Fig. \ref{comparando_c0} the total energies corresponding to the curves displayed in Fig. \ref{normaliza_h,c0}, recorded during the ZFC processes while heating the system up.
In Fig. \ref{comparando_c0}, we observe that the energy curves of the non-interacting case differ if using the same absolute field (a), but overlap if using the same reduced value $H/H_{A}=0.18$. Analogously, for the interacting case (and constant $H/H_{A}=0.18$), the curves differ if using the same absolute concentration (c) and overlap if using the same normalized value $c/c_{0}$. This result demonstrates the existence of an inner scale of energies as a function of $H_{A}$ and $c_{0}$.

\begin{figure}[!ht]
\includegraphics[angle = 0,width = 1.0\textwidth]{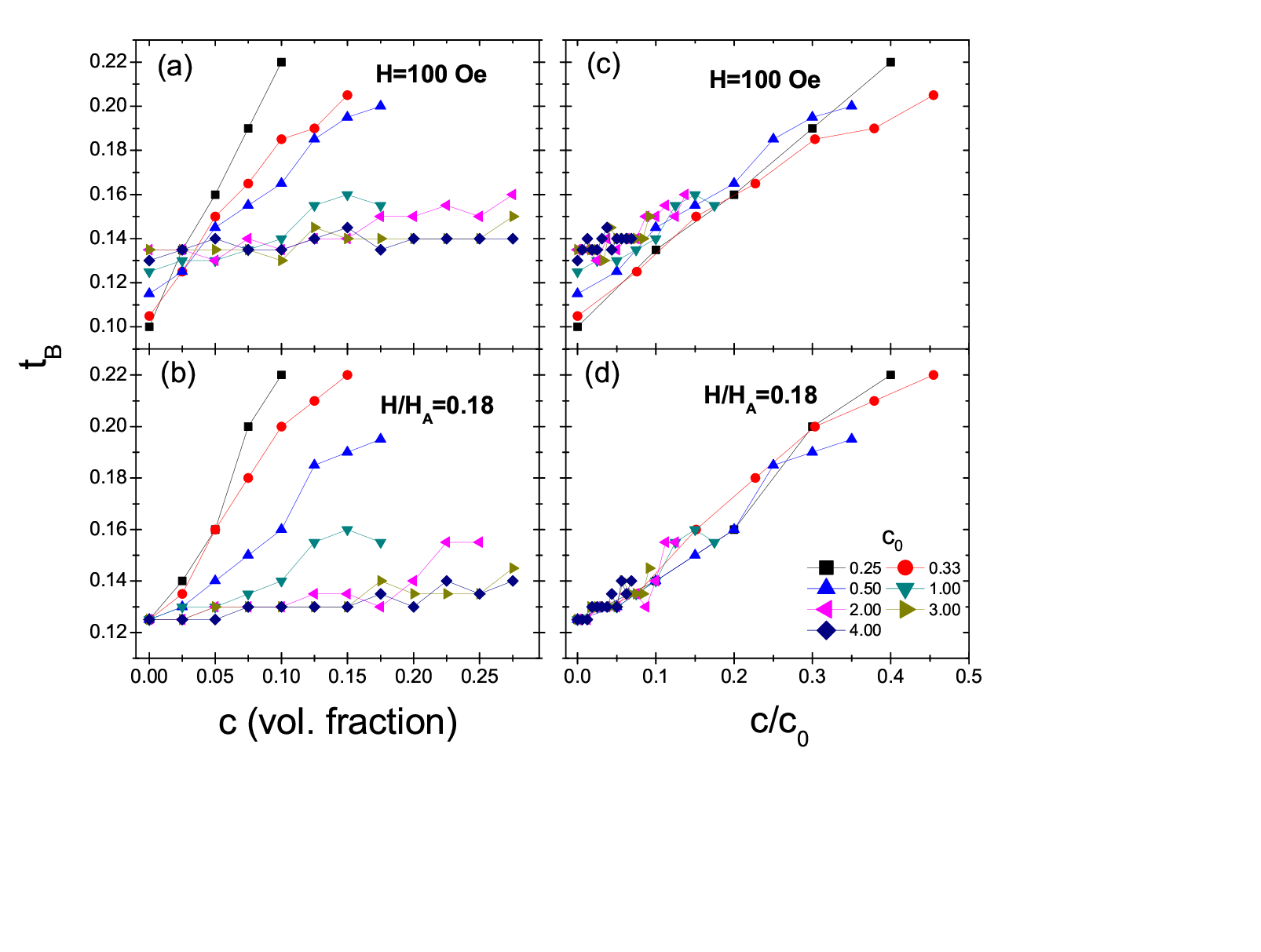}
\caption{(Color online) $t_{B}$ \textit{vs.} $c$ for, (a) $H=100~Oe$; (b) $H/H_{A}=0.18$; and \textit{vs.} $c/c_{0}$ for, (c) $H=100~Oe$; (d) $H/H_{A}=0.18$, for the different particle types of Table \ref{table-c0}.}\label{tB_cc0}
\end{figure}

We focus now on the $T_{B}$ ($t_{B}$ in reduced units) \textit{vs.} $c$ problem. Following the same procedure as in Fig. \ref{normaliza_h,c0}, we simulate ZFC processes for the different cases of Table \ref{table-c0} and systematically vary $c$, evaluating $t_{B}$ as the maxima of the curves. In Fig. \ref{tB_cc0}(a), we observe that the $t_{B}(c)$ data follows different trends under the same absolute values $H=100~Oe$ and $c=0.10$, but that all curves share however a common origin in the non-interacting $c=0.0$ case if using the same reduced field $H/H_{A}=0.18$ (Fig. \ref{tB_cc0}(b)). It is worth to emphasize the different trends observed in the $t_{B}$ \textit{vs.} $c$ data, which reproduce a rich variety of tendencies as mentioned in the introduction. A complete characterization of all different types would need however a more detailed analysis of the specific shapes of the curves, however this is beyond the scope of this work.

In Fig. \ref{tB_cc0}(c) it is shown that the apparently different trends of Fig. \ref{tB_cc0}(a) share however a more similar tendency if plotted \textit{vs.} $c/c_{0}$, even for the same absolute field $H=100~Oe$. Furthermore, all of them essentially overlap if using both the same $H/H_{A}=0.18$ value and plotted \textit{vs.} $c/c_{0}$, as shown in Fig. \ref{tB_cc0}(d). These results suggest (within the precision of the results) the existence of a general $t_{B}(H/H_{A};c/c_{0})$ curve, absorbing the different trends of Fig. \ref{tB_cc0}(a).

\begin{figure}[!ht]
\includegraphics[angle = 0,width = 1.0\textwidth]{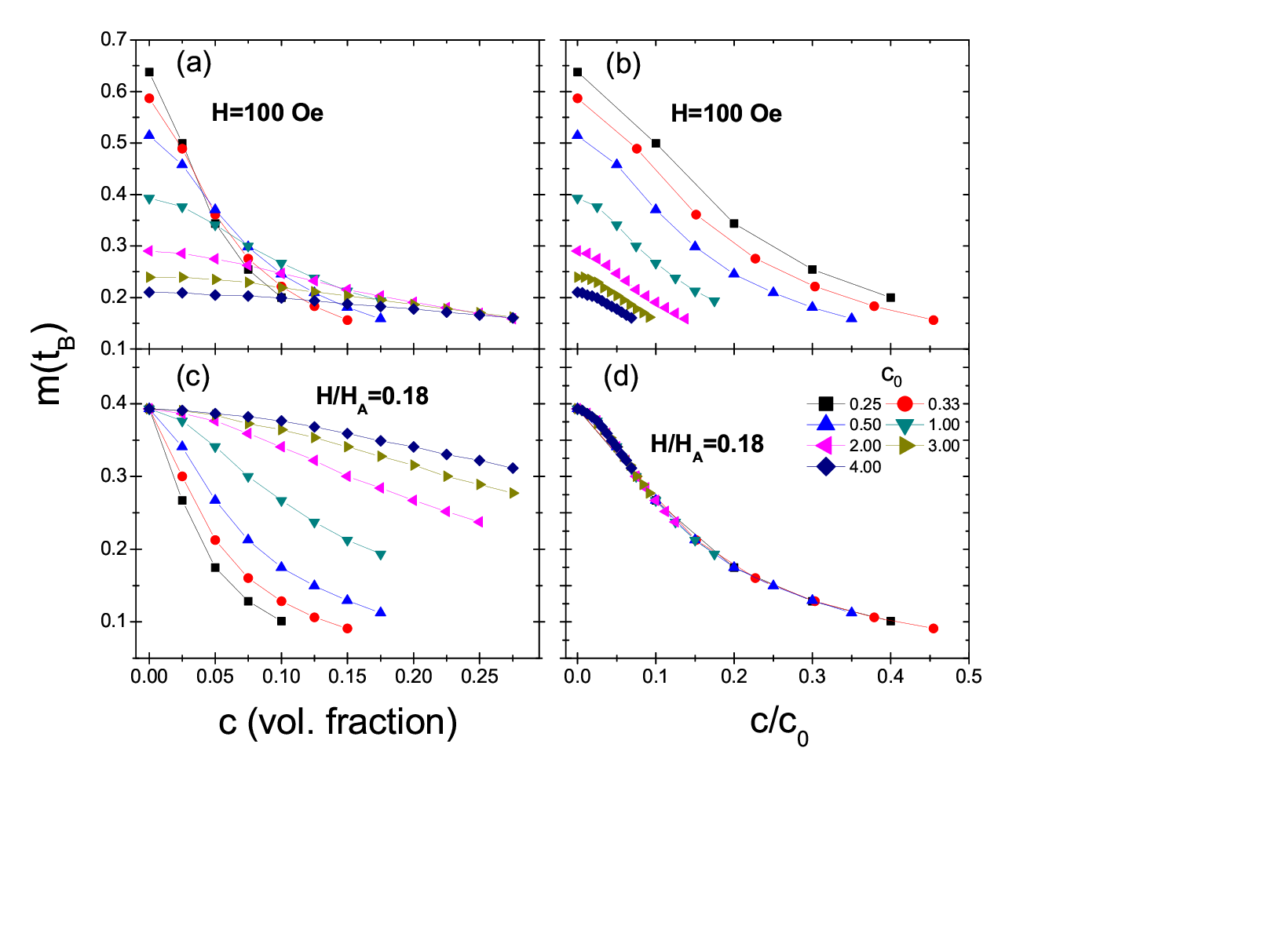}
\caption{(Color online) (a) Magnetization at $t_{B}$ \textit{vs.} $c$ for different $c_{0}$-values at a fixed field $H=100~Oe$; (b) same data as in (a), plotted \textit{vs.} $c/c_{0}$; (c) same data as in (a) but using the same reduced field $h=0.18$ for the different $c_{0}$ cases; (d) same values as in (c), plotted \textit{vs.} $c/c_{0}$.}\label{m_tB}
\end{figure}

It might be argued, however, that the precision of the $t_{B}(c)$ data is not enough to demonstrate the scaling of the SPM phenomenon, since there is some deviation between the $t_{B}(c/c_{0})$ curves. To erase such possible doubts we have followed a different approach: we analyzed the maxima of the curves at $t_{B}$, $m(t_{B})$, which is far more precise to evaluate than $t_{B}$. The results displayed in Fig. \ref{m_tB} demonstrate with high precision the role of the $c_{0}$ parameter as a scaling factor for the magnetic dipolar interaction: the $m(t_{B})$ data, which follows different trends if using absolute $H=100~Oe$ values, as shown in Figs. \ref{m_tB}(a),(b); or absolute $c=0.1$ values, as in Figs. \ref{m_tB}(a),(c); however perfectly overlaps in a common $m(t_{B})$ curve if using the same relative $H/H_{A}=0.18$ value and plotted \textit{vs.} $c/c_{0}$, as shown in Fig. \ref{m_tB}(d).

The main objective of the present work was to show theoretically the general character of the $2K/M^{2}_{S}$ parameter to scale the dipolar interaction strength, for the particular case of the ZFC/FC curves. For the direct connection with experimental measurements there are a few aspects that need to be taken into account. Firstly, that the results presented here are focused on uniaxial-anisotropy particles, size-monodisperse and homogeneously distributed in space; it is well known that other anisotropy contributions\cite{russier_jmmm2016}, aggregation\cite{abbasi_jpcC2011} and polydispersity\cite{zheng_jpcm2006} may significantly change the $T_{B}$ value. Secondly, the measurements need to be carried out under the same $H/H_{A}$ ratio, as discussed above.

Direct comparison with experiment at this stage is difficult because of the lack of a systematic study. Nevertheless, while we do not perform here a detailed analysis of experimental samples, we note that the usefulness of the $c_{0}$ parameter to correlate with experimental data is well proven. We have previously shown that it can be used to understand different experimental trends related to hyperthermia experiments (through the $M(H)$ hysteresis loops at different concentrations)\cite{martinez-boubeta_afm2012}: aparently different experimental trends describing the heating properties of nanoparticle systems as a function of the concentration, can in fact be described as particular regions of a general curve on scaling with respect to $c/c_{0}$. The new approach is its use to specifically understand ZFC/FC measurements. In particular our scaling analysis provides a framework for the analysis of experimental data in terms of the nature of physical arrangement of the particles. 

\section{Conclusions}

In summary, we report the existence of a general scaling factor for the influence of the dipolar interaction energy in single-domain entities. It is defined by the dimensionless parameter $c_{0}=2K/M^{2}_{S}$, and together with the anisotropy field $H_{A}$ serves to define a general behavior for such systems as governed by the dipolar and Zeeman energies. Specifically, we applied this parameter to show how the evolution of $T_{B}$ \textit{vs.} $c$ in assemblies of SPM nanoparticles, which may correspond to apparently very different trends, may in fact be scaled by this parameter. This approach may offer new tools to understand the apparently dissimilar tendencies reported in the literature, belong in fact to the same general $T_{B}(c/c_{0})$ curve. Remains as the subject of a future work the detailed analysis of the initial decrease in the $T_{B}$ \textit{vs.} $c$ curve reported in other works\cite{hansen_jmmm1998,abbasi_jpcC2011}. Finally, we note that the relevant parameter for magnetic hyperthermia, the hysteresis area (estimation of the heat that may be released), is directly related to the effective anisotropy of the system and consequently the blocking temperature as modulated by the dipolar interactions. Consequently the $T_B$ determined from FC/ZFC measurements, and its concentration scaling presents itself as a useful approach to the study of the effects of interactions in magnetic nanoparticle hyperthermia.

\section{Acknowledgements}
We acknowledge the Royal Society International Exchanges Scheme (IE160535). D.S. acknowledges the Xunta de Galicia for financial support under the I2C Plan.





\bibliographystyle{model1-num-names}
\bibliography{sample.bib}







\end{document}